\documentclass[aip,preprint]{revtex4-1}

\usepackage{graphicx}

\usepackage{amsmath}
\usepackage{amssymb}
\usepackage{amsthm}
\usepackage{hyperref}

\begin{document}

\title{Efficient and accurate description of adsorption in zeolites}

\author{Ji\v{r}\'{i} Klime\v{s}}
\email{klimes@karlov.mff.cuni.cz}
\affiliation{Department of Chemical Physics and Optics, Faculty of Mathematics and Physics, 
Charles University, Ke Karlovu 3, CZ-12116 Prague 2, Czech Republic}
\affiliation{J. Heyrovsk\'{y} Institute of Physical Chemistry, Academy of Sciences of the Czech Republic, 
Dolej\v{s}kova 3, CZ-18223 Prague 8, Czech Republic}
\author{David P. Tew}
\affiliation{Max-Planck-Institut f\"{u}r Festk\"{o}rperforschung, Heisenbergstra{\ss}e 1, 70569 Stuttgart, Germany}

\date{\today }

%-------------------------------------------------------------------------
\begin{abstract} 
%-------------------------------------------------------------------------

Accurate theoretical methods are needed to correctly describe adsorption on solid surfaces 
or in porous materials.
The random phase approximation with singles corrections scheme (RPA) and the second order M{\o}ller-Plesset
perturbation theory (MP2) are two schemes which offer high accuracy at affordable computational cost.
However, there is little knowledge about their applicability and reliability for different adsorbates and surfaces.
Here we calculate adsorption energies of seven different molecules in zeolite chabazite
to show that RPA with singles corrections is superior to MP2 not only in terms of accuracy 
but also in terms of computer time.
Therefore RPA with singles is suitable scheme for obtaining highly accurate adsorption energies in porous materials
and similar systems.

%-------------------------------------------------------------------------
\end{abstract}
%-------------------------------------------------------------------------

\maketitle

%------------------------------------------------------------------------- 
\section{Introduction}
%-------------------------------------------------------------------------

Adsorption of molecules on solid surfaces or in porous materials is a phenomenon
important for heterogeneous catalysis in industry, for (de)mineralization in nature, 
and for many other processes.
Computational modelling at atomic and molecular level greatly helps to identify
structures of adsorbates or to elucidate reactions catalyzed by solid surfaces.\cite{carrasco2009}
In particular, Kohn-Sham density functional theory (DFT)\cite{hohenberg1964,kohn1965} has been
an extremely useful tool for this task.
However, in many systems, substantial contributions to adsorption energies come from long-range
electron correlations (dispersion forces).\cite{tuma2006,hammerschmidt2012}
These are not accounted for by standard semi-local or hybrid DFT approximations and they can thus 
lead to large errors for adsorption energies.\cite{chakarova2006,sony2007,berland2009}
Adding dispersion corrections usually reduces the errors but they can remain large\cite{gautier2015} 
even for seemingly simple systems, such as semiconductors or oxides.\cite{alhamdani2015,zen2016} 
This means that the DFT based approaches are still far from being reliable enough to be used to obtain reference
quality adsorption energies.

To obtain highly accurate adsorption energies one needs a scheme that includes a high-level 
description of electron correlation effects.
Methods that are capable of this task are quantum Monte Carlo (QMC) techniques, such as diffusion Monte Carlo
or the coupled cluster scheme,
using at least singles, doubles, and perturbative triples excitations, CCSD(T).
For finite cluster calculations these methods have been shown to agree to within few per cent, 
see, e.g., the recent work in Refs.~\onlinecite{dubecky2014,alhamdani2017,brandenburg2019,dubecky2019}.
However, so far there has been less agreement between adsorption energies calculated within periodic 
boundary conditions,\cite{brandenburg2019} even though algorithmic improvements are likely to reduce
this issue in the future as well.\cite{zen2016qmc,hummel2017,zhang2019}
As an alternative, embedded cluster techniques can be used to calculate adsorption energies,\cite{lopez1999,valentin2002}
and the recent combination with low scaling coupled cluster implementations seems particularly promising.\cite{kubas2016}
Another alternative is to use less elaborate methods that could offer high accuracy nevertheless.
One of them is the second-order M{\o}ller-Plesset perturbation theory approximation (MP2) which 
has been widely used in quantum chemistry.
In fact, it has been also used to study adsorption using embedded finite cluster approach,
see, {\it e.g.} Ref.~\onlinecite{boese2016} for a recent work.
Moreover, periodic implementation of MP2 was used to study adsorption in zeolites.\cite{goltl2012MP2}
Another promising method available within periodic boundary conditions is the 
random phase approximation (RPA).\cite{harl2008,lu2009,harl2009,nguyen2009,olsen2011}
RPA surpasses MP2 in description of electron screening but it lacks second and higher order exchange effects,\cite{grueneis2009}
the relative importance of either of these is still rather unclear in general.\cite{ren2013}
The accuracy of RPA, even though it was shown to be consistent for several systems,\cite{harl2010,bleiziffer2012,torres2017}
was found to be too low to make it the method of choice. 
Several ways how to improve the accuracy of RPA have been explored, for example obtaining self-consistent RPA 
energies\cite{bleiziffer2012,bleiziffer2013} or using exchange-correlation kernels.\cite{olsen2013,olsen2014,bleiziffer2015}
Another approach that has lead to improved accuracy of RPA was to add the so-called singles corrections,
either the renormalized singles excitations (RSE)\cite{ren2011} or the $GW$ singles excitations (GWSE),\cite{klimes2015} 
that do not or do account for electron screening, respectively.
The high accuracy of RPA with singles has been demonstrated for systems such as molecular solids\cite{klimes2016,zen2018} 
and water adsorption on sodium chloride, hexagonal boron nitride, or graphene.\cite{klimes2015,alhamdani2017,brandenburg2019}
However, the studies on adsorption considered only a single adsorbate and substrate type at a time and more general
understanding of the accuracy and limitations of RPA with singles is missing.

In this work we want to understand the predictive power of MP2 and RPA with singles for the calculation
of adsorption energies in simple systems.
To achieve this goal we focus on adsorption of small molecules in zeolites, which are porous aluminosilicates.
Previous studies have shown that, rather surprisingly, many DFT based schemes give adsorption energies
 which are too large.\cite{goltl2011,bucko2013,shang2014,bucko2014}
This has even partly stimulated a development of improved dispersion correction methods.\cite{bucko2014}
In contrast, MP2 has been often shown to be close to CCSD(T) for finite cluster calculations of zeolites\cite{tuma2015}
and RPA with singles was shown to improve the description of CO adsorption on models of zeolites.\cite{rubes2018}
Due to the problems in describing adsorption in zeolites and the current availability of both RPA with singles and MP2
 it is interesting to ask: Which one of these is more reliable for the description of adsorption?
Is it the widely used MP2 or RPA which, until 2001,\cite{furche2001} has received almost no attention in quantum chemistry?
To answer these questions we calculated adsorption energies of seven molecules using MP2 and RPA with singles.
While they give results which are surprisingly close to each other for adsorption in bulk, RPA with singles
outperforms MP2 for adsorption on finite clusters.
Moreover, its computational cost is one order of magnitude smaller than that of MP2.
Our data, together with the previous results\cite{klimes2015,alhamdani2017,rubes2018,brandenburg2019} 
suggest that RPA with singles is the current method 
of choice for obtaining nearly reference quality adsorption energies in zeolites as well as other systems at affordable
computational cost.

%------------------------------------------------------------------------- 
\section{Systems}
%-------------------------------------------------------------------------

The basic tetrahedral binding motif of silica groups gives rise to a large number 
of different zeolite structures which can have unit cell volumes of several thousands of cubic {\AA}ngstr\"{o}ms.
MP2 and RPA are methods based on perturbation theory and, compared to simpler DFT approaches, 
have larger memory and computational time requirements.
Therefore, to obtain precisely converged adsorption energies, one needs
a system with relatively small simulation cell, below approximately 2000~\AA$^3$.
As was done in previous studies,\cite{goltl2011,goltl2012MP2} we picked the chabazite structure
with unit cell composition Si$_{11}$AlO$_{24}$H.

\begin{figure}
       \includegraphics[width=16cm,clip=true]{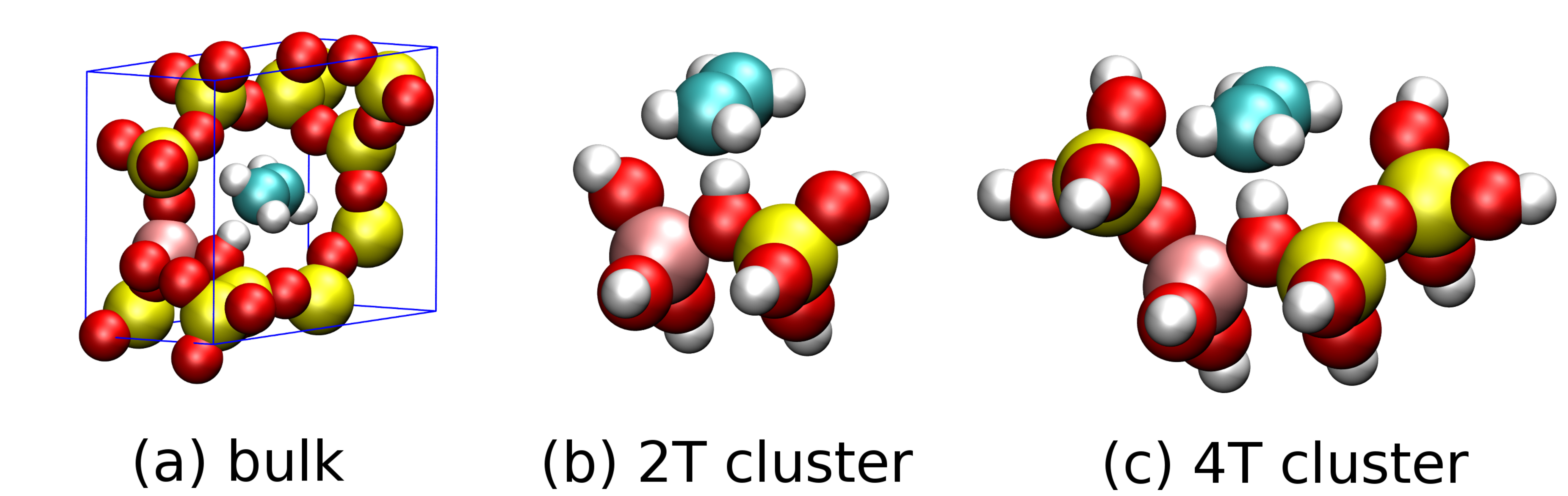}
   \caption{The unit cell of chabazite with adsorbed ethylene molecule (a) and, in (b) and (c), the 2T and 4T models
used to perform finite cluster calculations.}
\label{fig:strucs}
\end{figure}

We calculated adsorption energies of seven molecules:
methane, ethane, ethylene, acetylene, propane, CO$_2$, and H$_2$O.
To obtain the required geometries we first optimized the chabazite with adsorbed 
methane using the optB88-vdW functional.\cite{dion2004,soler2009,klimes2010} 
The zeolite framework was held fixed in all the subsequent calculations.
This was simply to reduce the number of calculations needed to obtain the results. 
Moreover, this is not an issue as our primary interest here is to understand the differences 
between  MP2 and RPA adsorption energies.
Structural optimization, in some cases combined with molecular dynamics, was used to obtain the 
adsorption structures of the other molecules.
The unit cell of chabazite with adsorbed ethylene is shown in Fig.~\ref{fig:strucs}(a).

Furthermore, we have created two finite clusters, one with 2 tetrahedral sites (2T, AlSiO$_7$H$_7$) 
and one with four tetrahedral sites (4T, AlSi$_3$O$_{13}$H$_{11}$) to assess the quality
of the adsorption energies using reference quality method.
For the clusters, the broken bonds were capped with hydrogens which were subsequently relaxed 
keeping the positions of all the other atoms fixed.
The clusters that we used together with adsorbed ethylene are shown in Fig.~\ref{fig:strucs}(b)
for the 2T cluster and in panel (c) of the figure for the 4T clusters.
All the structures are given in the SI.\cite{supplementary}

%------------------------------------------------------------------------ 
\section{Computational setup}
%------------------------------------------------------------------------

The adsorption energies $E^{\rm ads}$ were obtained as $E^{\rm ads}=E_{\rm zeo+mol}-E_{\rm zeo}-E_{\rm mol}$,
where $E_{\rm zeo+mol}$,  $E_{\rm zeo}$, and  $E_{\rm mol}$ are the energies of zeolite (either bulk or finite cluster)
with adsorbed molecule, the bare zeolite, and the isolated molecule, respectively.
In all three cases, the geometries of the molecule or zeolite are identical, 
that is we are calculating interaction energy.
Moreover, all three calculations used the same simulation cell.
This reduces errors due to k-point sampling and basis-set incompleteness.
For the bulk calculations, there is a small residual error due to interaction with periodic images for the molecule, 
this amounts to 0.5~kJ/mol for water and is negligible in other cases.
In any case this does not affect our comparison as the error is similar for different methods.

The calculations employing periodic boundary conditions were performed using 
the VASP program.\cite{kresse1993,kresse1996}
VASP uses a plane-wave basis set and the projector-augmented wave (PAW) 
ansatz for the wavefunction.\cite{blochl1994,kresse1999}
The MP2 calculations were performed using the standard implementation in VASP.\cite{marsman2009,grueneis2010} 
While the recent $O(N^4)$ implementation of MP2 was shown to be more efficient for the systems considered
here,\cite{schafer2017} we did not use it as the corrections for the Gamma-point singularity
of the Coulomb interaction were not yet available.\cite{gajdos2006,liao2016}
We used the cubic scaling code for the RPA and singles calculations.\cite{kaltak2014rpa1,kaltak2014rpa2,klimes2015}
Two sets of PAW potentials were used, first ``standard'' PAWs for testing and volume convergence for the finite clusters, 
and, second, ``hard'' PAWs which treat also semi-core electrons as valence and are almost norm-conserving.\cite{klimes2014NC}
The specific PAW potentials that we used are listed in the SI.\cite{supplementary}

The total RPA energy is composed of the exact exchange (EXX) and random phase correlation (RPAc) energy components.
The singles corrections can be added either in the RSE, or in the GWSE flavor.
The Perdew-Burke-Ernzerhof (PBE)\cite{perdew1996} functional was used as a starting point for the RPA calculations.
The EXX and RSE adsorption energies converge quickly with the basis set size and we have used a 1050~eV 
basis-set cut-off to obtain their values. 
This cut-off guarantees a convergence of the adsorption energies to within 0.05~kJ/mol, 
the individual energies are provided in the SI.\cite{supplementary}
The RPAc and GWSE energies have a stronger dependence on the basis-set cut-off due to the two-electron cusp.
We obtained the energies for several values of cut-offs and used data for cut-offs  850, 950, and 1050~eV to 
extrapolate to infinite basis-set size limit. 
In the case of GWSE, we extrapolated the difference between GWSE and RSE corrections.
In the extrapolation, we assumed that the basis-set incompleteness error is proportional to $E_{\rm cut}^{-3/2}$.
The auxiliary basis-set energy cut-off for response related properties ({\tt ENCUTGW} tag for VASP) was set to one half 
of the basis-set cut-off ({\tt ENCUT}).\cite{harl2008}
We note that the adsorption energy converges rather quickly with the basis-set size.
Interestingly, this holds both for the adsorption energies obtained using a given combination of {\tt ENCUT} and {\tt ENCUTGW}
and adsorption energies obtained for a given value of {\tt ENCUT} and extrapolated with the auxiliary basis set {\tt ENCUTGW} 
in a single calculation.\cite{harl2008}
This is likely because the change of density upon adsorption is small and the leading contributions to
basis-set incompleteness errors for the adsorbed system and the individual parts cancel each other.\cite{gulans_unp,klimes2014NC}
Using a basis-set cut-off of 750~eV and the {\tt ENCUTGW} extrapolation 
would lead to errors of less than 1.3\% or 0.4~kJ/mol in the adsorption energies, 
which is acceptable in most of applications.

The total MP2 energy is composed of the Hartree-Fock energy (HF) and the MP2 correlation energy, which
itself can be divided into direct MP2 (dMP2) and exchange MP2 (xMP2) parts.
The HF part was obtained using a basis-set cut-off of 1050~eV.
The MP2 implementation has higher computational requirements compared to the RPA implementation
and we were able to obtain data only for basis-set cut-offs of 650~eV and, for bulk only, 750~eV.
The infinite basis-set limit of the xMP2 adsorption energies was obtained by extrapolating data calculated 
with 650~eV basis-set cut-off and cut-offs of 325 and 425~eV for the auxiliary basis set ({\tt ENCUTGW}).
For the bulk, the results are within 0.05~kJ/mol of data for which the second point for extrapolation used 
750~eV cut-off for the basis-set and 375~eV cut-off for the auxiliary basis.
One exception is CO$_2$ where the difference is 0.14~kJ/mol, which is still small not to affect our conclusions.
We note that the xMP2 contribution to energy is below 1~kJ/mol for the hydrocarbons while for CO$_2$ and H$_2$O
the contribution is larger, being approximately $-3$ and $-4$~kJ/mol, respectively.
%These data were extrapolated to the complete basis set limit to obtain the xMP2 energy.
%
For the dMP2 part, we obtained additional data using the RPA code and extrapolated to the complete
basis set limit using cut-off values of 850, 950, and 1050~eV.
As for the RPA calculations, the auxiliary basis-set cut-off energy was set to one half of the cut-off for the orbitals.
In the bulk calculations of EXX and HF energies, a 2$\times$2$\times$2 k-point set was used, 
for all the other calculations (singles corrections and RPA and MP2 correlation energies) only the gamma point was used 
for k-point sampling.

Interaction energies on a finite cluster were obtained using VASP for RPA, RSE, and MP2 and using Turbomole, 
which provided basis set limit CCSD(T) reference and MP2 energies for comparison.
In the VASP calculations, the HF and EXX energies were obtained using a 1050~eV basis-set cut-off 
and a cell with at least 18~\AA~side.
The RPA and MP2 binding energies were obtained by a composite procedure.
First, a basis-set converged data were obtained in a cubic cell with 9~\AA~side, using the same settings
for basis-set cut-offs as in bulk.
Second, correction to an infinite cell volume was added to obtain data converged with both
the basis and cell volume.
The finite size correction was calculated as the difference between the binding energy in a 9~\AA\ cell 
and binding energy extrapolated to infinite volume using cells with up to 12~\AA~side.
Similar correction was used to obtain the infinite cell limit of the RSE correction.

The Turbomole calculations used the explicitly correlated coupled cluster approach 
CCSD(T)(F12*)\cite{hattig2010} and the cc-pVTZ-F12 basis set.\cite{peterson2008}
The Slater-type correlation factor was used with an exponent of 1.0 a$_0^{-1}$, 
together with the specially optimised RI basis sets.\cite{yousaf2008}
We also obtained MP2 energies using the MP2-F12 approach\cite{bachorz2011} and cc-pVTZ-F12 
basis set to provide reference data to compare to VASP. 
There is a satisfactory agreement between the MP2 adsorption energies obtained with VASP 
and Turbomole, the average absolute difference is only 0.1 kJ/mol. 
Canonical coupled cluster calculations were possible for all cases except for the 4T clusters, 
which were performed using the PNO-CCSD(T0)(F12*) approach.\cite{schmitz2014,schmitz2016}
We found that a very tight value of $T_\text{pno}$ parameter is needed to obtain converged 
adsorption energies and we used $T_\text{pno} = 10^{-8}$. 
To obtain the full contribution of triples, we scaled the T0 value by the T/T0 ratio obtained 
using the cc-VDZ-F12 basis set, which was possible to compute for all 4T clusters.

%------------------------------------------------------------------------- 
\section{Results}
%------------------------------------------------------------------------
\label{sec:results}

%------------------------------------------------------------------------- 
\subsection{Adsorption in bulk material}
\label{sec:res:bulk}
%------------------------------------------------------------------------- 

\begin{figure}
       \includegraphics[width=8cm,clip=true]{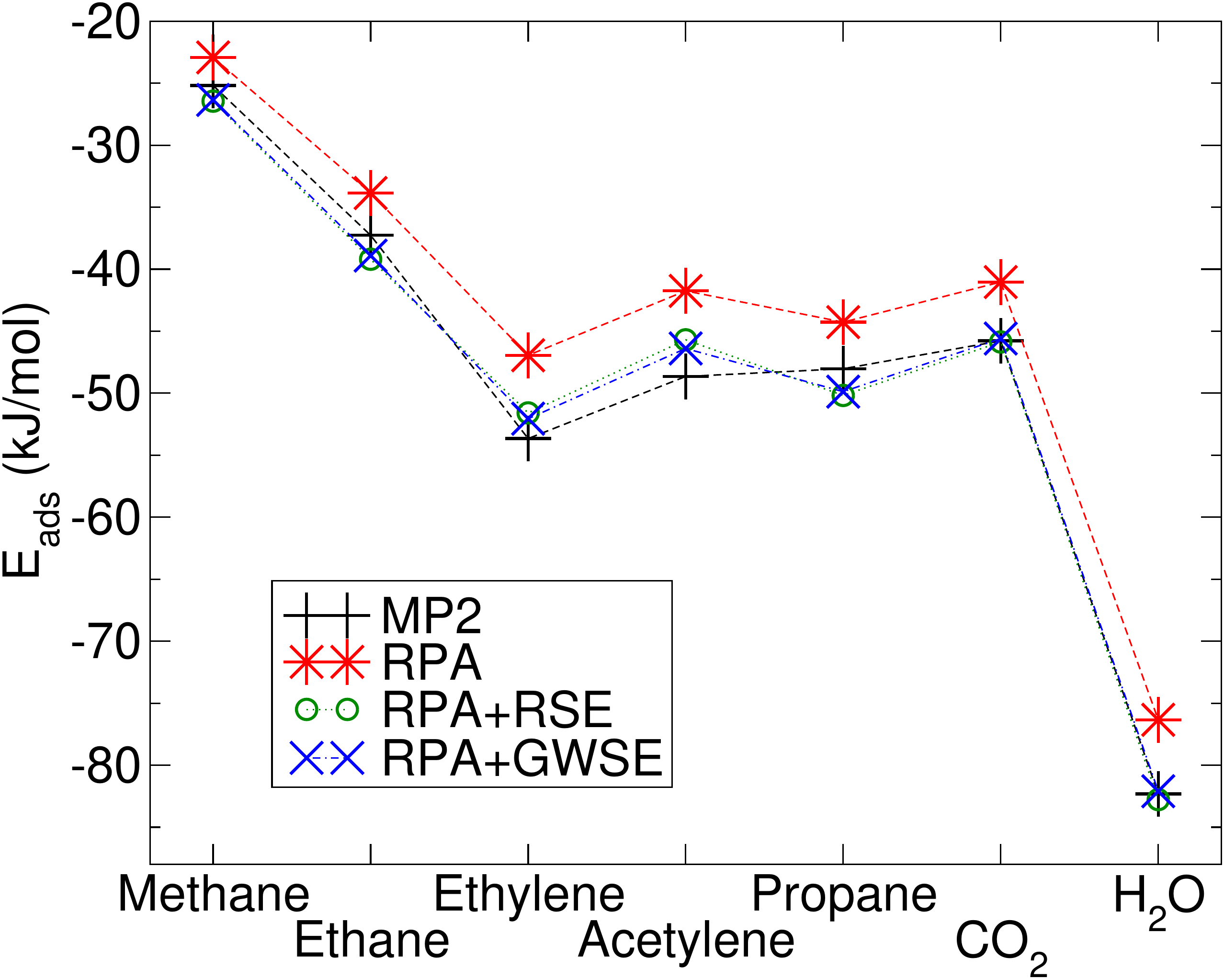}
   \caption{Adsorption energies of different molecules in chabazite
for HF, MP2, and RPA-based methods.}
\label{fig:ads_bulk}
\end{figure}

The adsorption energies of the different molecules in bulk chabazite are summarized 
in Table~\ref{tab:ads_bulk} and plotted in Fig.~\ref{fig:ads_bulk}.
Looking at the graph, the first striking thing is the close agreement between 
MP2 and RPA with singles corrections, either RSE or GWSE.
As the table shows, MP2 and RPA+GWSE adsorption energies differ only by 0.6~kJ/mol and 0.7~kJ/mol
for water and carbon dioxide, respectively.
The differences are larger for other systems, but they are still close to one or two kJ/mol
or few per cent.
Given the differences between RPA with singles and MP2, the close agreement points to a fact that
they provide highly accurate adsorption energies.
Moreover, the agreement between RPA with singles and MP2 is unexpected as the results previously 
published by G\"{o}ltl {\it et al.} imply that there should be a much larger difference, {\it e.g.} 
at least 8~kJ/mol for propane.\cite{goltl2012MP2}
In passing, we note that for the systems considered here the computational cost of RPA with GWSE corrections is about
an order of magnitude smaller than the cost of MP2 and the memory requirements are few times smaller as well.

\begin{table}
\caption{Adsorption energies of different molecules with bulk chabazite obtained 
for different methods. Data are in kJ/mol}
\label{tab:ads_bulk}
\begin{ruledtabular}
\begin{tabular}{lccccccc}
System   & HF    & xMP2 & MP2        & EXX          &RPA     & +RSE   & +GWSE  \\
\hline                  
Methane  &$4.6$  &$0.1$  &$-25.6$   & 10.1   &$-23.0$ &$-26.5$ &$-26.4$ \\
Ethane   &$10.2$ &$0.3$  &$-37.0$   & 18.5   &$-33.1$ &$-38.5$ &$-38.2$ \\
Ethylene &$-3.2$ &$-0.9$ &$-54.2$   & 3.8    &$-46.9$ &$-51.6$ &$-52.1$ \\
Acetylene&$-6.4$ &$-0.3$ &$-49.1$   & $-$0.2 &$-41.7$ &$-45.6$ &$-46.3$ \\
Propane  &$10.4$ &$0.4$  &$-47.6$   & 19.5   &$-43.2$ &$-49.1$ &$-48.8$ \\
CO$_2$   &$-10.1$ &$-2.9$ &$-46.3$   & $-$2.6 &$-41.0$ &$-45.9$ &$-45.6$ \\
H$_2$O   &$-54.4$ &$-3.8$ &$-82.7$   & $-$45.0&$-76.4$ &$-82.8$ &$-82.1$ \\
\end{tabular}
\end{ruledtabular}
\end{table}

The differences between the RSE and GWSE corrections are small,
typically below one per cent so that the data almost overlap in Fig.~\ref{fig:ads_bulk}.
Looking at the results in a more detail, two cases emerge, for the first, adding GWSE 
over RSE leads to a small reduction of binding.
This is the case of ethane, propane, CO$_2$, and water, where the binding is reduced by 0.3~kJ/mol
for the first three and by 0.8~kJ/mol for water.
For ethylene and acetylene, using GWSE increases the RPA+RSE binding by 0.5~kJ/mol and by 0.7~kJ/mol, respectively.
We note that increase of binding upon adding the GWSE corrections over RSE was also observed 
for molecular solids where the constituent molecules contained delocalized $\pi$ electrons.\cite{klimes2016}
For such systems screening is important and the GWSE corrections are expected to be sizeable.\cite{klimes2015}

\begin{figure}
       \includegraphics[width=8cm,clip=true]{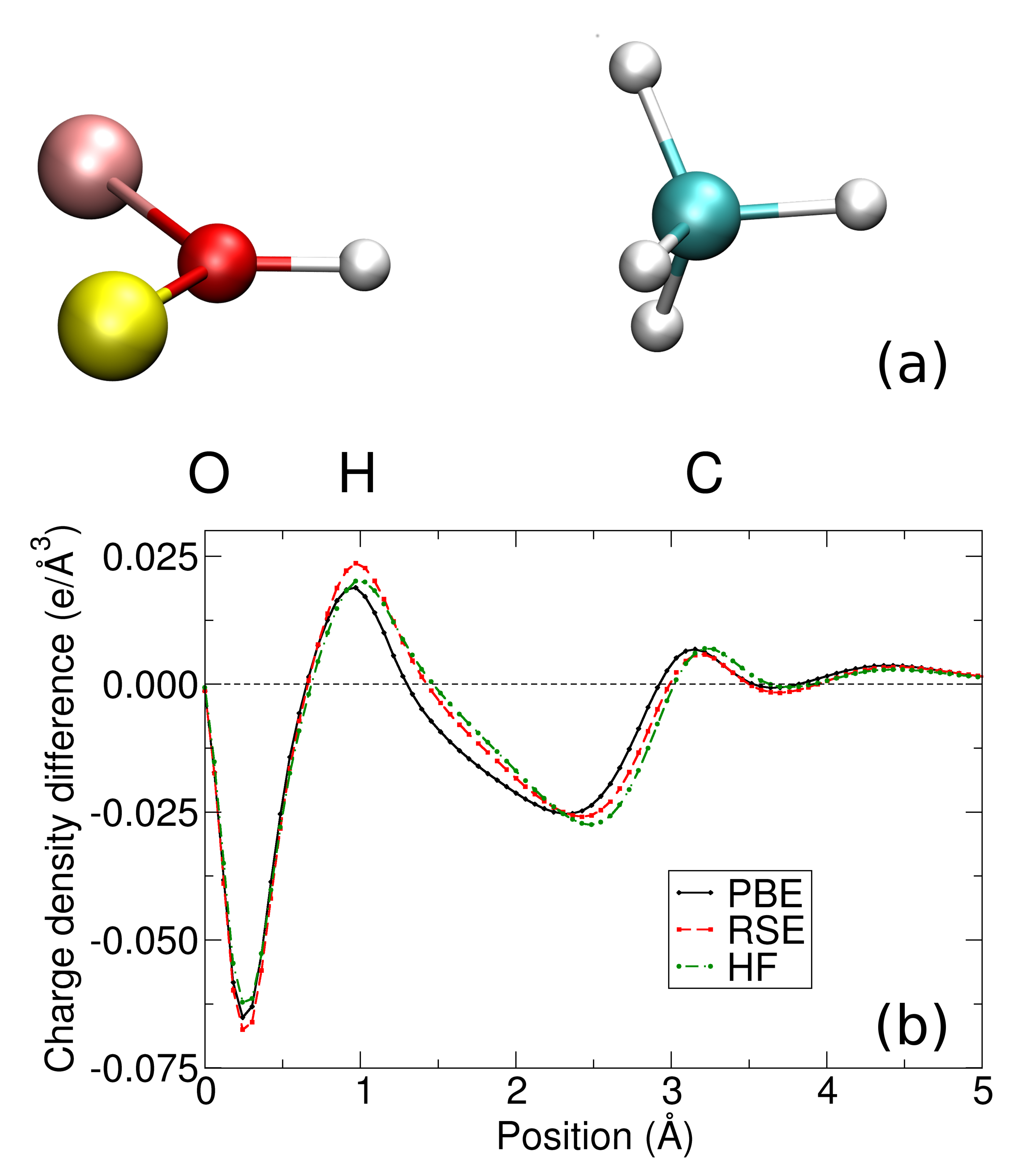}
   \caption{The structure of adsorbed methane showing part of the framework (a) and the charge density 
            difference upon adsorption along the direction of the OH bond (b).}
\label{fig:rho_diff}
\end{figure}

The physical reason why the singles corrections increase the binding is that they
change the density matrix towards the HF state,\cite{klimes2015} which is known to bind stronger compared
to EXX based on PBE orbitals.\cite{ren2011}
Other properties, such as the electron density, are affected as well upon adding the singles.
For example, we have recently shown that the charge density difference ($\Delta\rho$) for noble gas dimers obtained with singles corrections 
is close to HF and coupled cluster density differences in the bonding region.\cite{modrzejewski2019}
We observe similar changes in the density differences for the systems studied here.
Specifically, in Fig.~\ref{fig:rho_diff} we show the charge density difference for adsorption of methane,
the line profile runs along the direction of the O-H bond.
Clearly, the charge accumulation (positive values) and charge depletion regions as given by PBE are shifted
in the direction of the acidic hydrogen (small values of $x$) compared to HF.
When the singles corrections are added to PBE, $\Delta\rho$ becomes closer to the HF result.
This is consistent with the results observed for the neon dimer.\cite{modrzejewski2019}

The RPA is known to give adsorption or binding energies which are too 
weak.\cite{ren2011,eshuis2012,bleiziffer2012,olsen2013RPA,olsen2014,klimes2015,alhamdani2017}
This holds also here if we consider the MP2 or RPA+GWSE values as the reference.
The specific values of the GWSE corrections, that can be taken as a rough estimate of the
RPA error, are then between 3.5 and 5.8~kJ/mol.
These values amount to around 10\% of the adsorption energy, which is consistent with typical errors
of RPA binding energies.
%

%compare to Goltl and Piccini, discuss error of normal PAWs
Let us now compare our results to those presented previously in the literature.
G\"{o}ltl and coworkers\cite{goltl2012MP2} have obtained RPA and MP2 adsorption energies 
for methane, ethane, and propane in chabazite for four different positions of the acidic hydrogen.
Moreover, Piccini~{\it et al.} have obtained adsorption energies for the same molecules using 
the MP2:PBE-D approach.\cite{tuma2006,piccini2015jpcc}
The RPA adsorption energies of methane, ethane, and propane obtained by G\"{o}ltl and coworkers 
($-$19.69, $-$27.65, and $-$34.96~kJ/mol, respectively) 
are several kJ/mol smaller in magnitude compared to the values obtained here.
Note that our structures have the acidic hydrogen bound to the O(1) oxygen of chabazite and we compare
to the data of G\"{o}ltl~{\it et al.} for the same position.
We have found that part of the difference could be attributed to the PAW potentials used in Ref.~\onlinecite{goltl2012MP2}.
G\"{o}ltl and coworkers used ``standard'' PAW potentials  while we have used ``hard'' PAW potentials 
with small core radii.\cite{klimes2014NC} 
We find that the ``hard'' potentials give binding which is 0.5~to 1~kJ/mol stronger compared to the ``standard'' ones
for all the systems considered here.
This difference is significant when comparing the quality of different methods as well as when comparison to experimental
data is made.
The use of hard PAWs or pseudopotentials could be especially important when trying to understand difficult systems, 
such as the structure of adsorbates in zeolites under high pressure conditions.\cite{arletti2017,fischer2019pcm}
Turning to MP2, our adsorption energy for methane agrees well with the value of $-25.13$~kJ/mol reported by G\"{o}ltl~{\it et al.}
for the O(1) structure.
Piccini~{\it et al.} gives a similar value ($-$25.34~kJ/mol), however, in their structure hydrogen
is bound to the O(4) oxygen.
(We note that in Ref.~\onlinecite{piccini2015jpcc} a comparison is made to G\"{o}ltl's data for the O(3) structure.)
G\"{o}ltl and coworkers found that the structures where the acidic hydrogen is bound to  oxygen O(1), as used here, 
and to oxygen O(4), as used by Piccini~{\it et al.}, give adsorption energies which are larger by 0.25 and 1.32~kJ/mol
for ethane and propane, respectively.
Our MP2 adsorption energies are larger by 0.8~kJ/mol for ethane and by 3.5~kJ/mol for propane compared to those 
of Piccini~{\it et al.} and part of this difference could be attributed to the different adsorption site.
Given the differences in computational set-up and adsorption structures, the remaining difference is reasonably small 
and we consider our data and the data of Piccini~and co-workers to be in a good agreement.
The MP2 adsorption energies of G\"{o}ltl and coworkers for ethane and propane, $-40.68$ and $-55.49$~kJ/mol
are then too large compared to our or Piccini's data.
%

%RPA
%us O(1)    -23.0  -33.1  -43.2 
%goltl O(1) -19.69 -27.65 -34.96
%with MD
%goltl       -15.5 -19.7 -28.5

%MP2
%us O(1)      -25.6  -37.0  -47.6 
%goltl O(1)   -25.13 -40.68 -55.49
%goltl O(4)   -28.24 -40.43 -54.17
%Piccini O(4?) -25.34 -36.21 -44.13  

The close agreement between RPA with singles and MP2 calls for a more detailed comparison of the results
which reveals clear differences, visible already in Fig~\ref{fig:ads_bulk}.
Specifically, alkanes (methane, ethane, propane) are bound less by MP2 
than by RPA with singles (either RSE or GWSE).
In contrast, the MP2 binding is stronger for ethylene and acetylene, by 2.1 and 2.8~kJ/mol, respectively, 
when comparing to RPA+GWSE.
Comparing the different schemes, MP2 includes the second-order exchange contribution to the correlation energy, 
the xMP2 term, which is not accounted for in RPA.
In contrast, RPA sums a class of terms up to infinite order and accounts for screening.
The lack of screening leads to too large interaction energies for systems with delocalized electrons, 
and this would help to explain the larger MP2 adsorption energies for ethylene and acetylene.
The xMP2 term is expected to contribute the most to the adsorption energy when there is a significant overlap 
between the fragments.
Consequently, the xMP2 contributions to the adsorption energies for the hydrocarbons are small, 
less than 1~kJ/mol, see Table~\ref{tab:ads_bulk}.
Only for CO$_2$ and water the xMP2 contributions are larger, being $-$2.9 and $-$3.8~kJ/mol, respectively.
For all the systems, the difference between MP2 and RPA+GWSE adsorption energies 
has the same sign as the xMP2 contribution and adding terms such as the second-order screened exchange\cite{grueneis2009,hummel2019}
could further improve the accuracy of RPA.
We reiterate that the original goal of this work was to understand how well RPA with singles performs
compared to MP2 which we expected to be accurate enough to provide reference data for adsorption energies in zeolites.
This is based on the small corrections between MP2 and CCSD(T) for finite clusters\cite{piccini2015jpcc,tuma2015,rubes2017} 
and also on previous results of G\"{o}ltl and coworkers.\cite{goltl2012MP2}
With the current results at hand, the question then arises, if the RPA with singles method is actually 
more accurate than the (more computationally demanding) MP2 scheme. 
Given the small energy differences involved, very precise reference data are needed and such
can't be currently obtained for the periodic material.
Therefore, we used finite cluster models of the adsorption site and we discuss the results now.

%------------------------------------------------------------------------- 
\subsection{Adsorption on clusters}
\label{sec:res:clus}
%------------------------------------------------------------------------- 

\begin{figure}
       \includegraphics[width=8cm,clip=true]{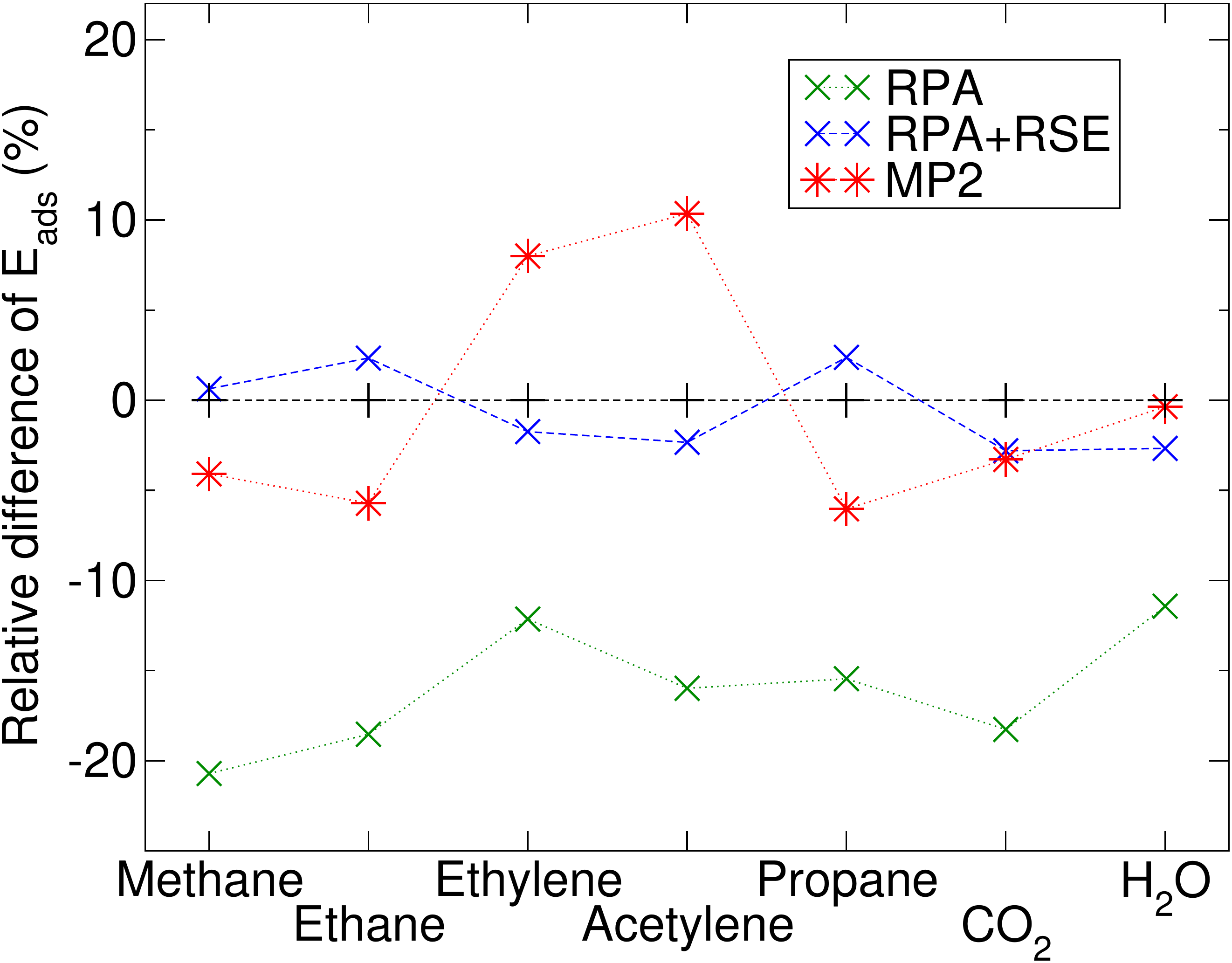}
   \caption{Relative differences of the adsorption energies on the 
2T cluster with respect to the CCSD(T) reference data.}
\label{fig:ads_clus1}
\end{figure}

We now discuss the results obtained for the finite clusters, starting with the 
smaller 2T one.
Table~\ref{tab:ads_clus1} summarizes the adsorption energies, relative differences 
from the CCSD(T) reference data are presented in Fig.~\ref{fig:ads_clus1}.
One can see that in all but one case the RPA+RSE adsorption
energies are closer to the CCSD(T) reference than are the MP2 adsorption energies.
Not only that, the results of RPA+RSE are rather consistent, the relative errors are
within 3\% for all the systems.
In contrast, MP2 gives almost no error for water while the error reaches 10\%
for ethylene and acetylene.
In agreement with previous results, the adsorption energies predicted by RPA 
are underestimated by 10 to 20\%.

\begin{table}
\caption{Adsorption energies of different molecules on the 2T cluster
as obtained using RPA, RPA+RSE, MP2, and CCSD(T) which serves as a reference.
All the data are in kJ/mol.}
\label{tab:ads_clus1}
\begin{ruledtabular}
\begin{tabular}{lcccc}
System     & RPA    & +RSE   & MP2      & CCSD(T) \\
\hline
Methane    &$-$8.7  &$-$11.1 &$-$10.6   &$-$11.0 \\
Ethane     &$-$9.4  &$-$11.8 &$-$10.9   &$-$11.6 \\
Ethylene   &$-$18.7 &$-$20.9 &$-$23.0   &$-$21.3 \\
Acetylene  &$-$14.0 &$-$16.3 &$-$18.4   &$-$16.7 \\
Propane    &$-$10.0 &$-$12.1 &$-$11.1   &$-$11.8 \\
CO$_2$     &$-$12.0 &$-$14.2 &$-$14.1   &$-$14.6 \\
H$_2$O     &$-$39.8 &$-$43.7 &$-$44.7   &$-$44.9 \\
\end{tabular}
\end{ruledtabular}
\end{table}

Finally, we consider adsorption on the large 4T cluster for which the results
are summarized in Table~\ref{tab:ads_clus2}.
Comparing first MP2 with RPA+RSE, one can make similar observations as from 
the data for bulk and the 2T cluster.
Specifically, alkanes are bound by about 1~kJ/mol less by MP2, 
ethylene and acetylene are bound by about 3~kJ/mol or 10\% more by MP2, 
the binding is within 0.1~kJ/mol for CO$_2$.
MP2 predicts 1~kJ/mol stronger binding for water compared to the RPA+RSE, this is similar
to the 2T cluster, but opposite to the binding in bulk.
Table~\ref{tab:ads_clus2} also lists the PNO-CCSD(T)(F12*) data for reference.
The data show overbinding of ethylene and acetylene by MP2.
The alkanes are bound too strongly by RPA+RSE, by about 1.5~kJ/mol which is more than for
the smaller 2T cluster where the difference was at most 0.3~kJ/mol for propane.
A similar increase in the binding strength for the alkanes can be observed for MP2.
MP2 gave about 0.5~kJ/mol too weak adsorption energies on the 2T cluster
while for the 4T cluster the binding is about 0.5~kJ/mol too strong.

\begin{table}
\caption{Adsorption energies of different molecules on the 4T cluster as calculated
with RPA, RPA+RSE, MP2, and the reference PNO-CCSD(T)(F12*). 
All the data are in kJ/mol.}
\label{tab:ads_clus2}
\begin{ruledtabular}
\begin{tabular}{lcccc}
System & RPA & +RSE & MP2& PNO-CCSD(T) \\
\hline
Methane &$-12.8$ &$-16.2$ &$-15.3$ &$-14.7$ \\
Ethane  &$-15.1$ &$-18.8$ &$-17.9$ &$-17.4$ \\
Ethylene  &$-30.7$ &$-35.3$ &$-38.1$ &$-32.7$ \\
Acetylene  &$-19.1$ &$-22.7$ &$-25.5$ &$-21.8$ \\
Propane &$-16.4$ &$-19.9$ &$-19.0$ &$-18.2$ \\
CO$_2$  &$-23.9$ &$-27.7$ &$-27.7$ &$-28.9$ \\
H$_2$O  &$-64.7$ &$-71.0$ &$-72.0$ &$-72.0$ \\
\end{tabular}
\end{ruledtabular}
\end{table}

\subsection{Comparison to density functional theory methods}

We now use the adsorption energies calculated with RPA with singles scheme to assess 
the predictions made by dispersion corrected DFT functionals.
It is well known that adsorption energies in zeolites predicted by current DFT approximations 
are not very accurate and are typically substantially 
overestimated.\cite{goltl2012MP2,thang2014,shang2014,piccini2015jpcc,rubes2018,rocca2019prm,piccini2019}
One exception is the DFT/CC scheme proposed by Bludsk\'{y} and coworkers.\cite{bludsky2008}
Here, we have obtained adsorption energies using the PBE-TS/IH method of Bu\v{c}ko and coworkers,\cite{bucko2013}
and the PBE-D3 and, for the finite cluster only, the PBE-D4 approaches of Grimme and coworkers.\cite{grimme2010,grimme2011damp,caldeweyher2017}

The adsorption energies are collected in Table~\ref{tab:ads_dft} for bulk and in Table~\ref{tab:ads_dft_2T}
for the 2T cluster.
The PBE-D3 scheme substantially overestimates the adsorption energies of hydrocarbons, by around 16~kJ/mol 
for propane in bulk and by up to 3.4~kJ/mol for acetylene on the 2T cluster.
Interestingly, the data for the 2T cluster show that the recent D4 correction does not bring 
any significant improvement compared to the D3 scheme.
Moreover, while the use of hybrid functionals has been found to improve the description of structural properties 
of zeolites,\cite{goltl2012structure,cutini2019} using the hybrid PBE0 functional does not 
improve the results substantially over those obtained with PBE.
The only molecule where PBE-D3 underestimates the binding is CO$_2$ and it's worth mentioning that
similar unexpected underbinding was also observed for CO$_2$ crystal.\cite{klimes2016}
The PBE-TS/IH method was shown to improve adsorption energies for Cs-exchanged chabazite.\cite{shang2014,bucko2014}
While PBE-TS/IH performs reasonably well for ethylene and acetylene, it considerably overestimates
the adsorption energies of alkanes.
The case of CO$_2$ becomes even more puzzling as PBE-TS/IH underestimates its adsorption energy 
by some 6~kJ/mol for bulk and by $2.1$ kJ/mol for the 2T cluster.
Overall, the results illustrate one of the benefits of RPA over dispersion corrected DFT 
which is the consistency of RPA errors.
Finally, we note that ``standard'' PAW potentials give interaction energies which are artificially too strong 
for CO$_2$ and H$_2$O, by 0.9 and 1.5~kJ/mol, respectively, for alkanes the differences to the more precise 
``hard'' PAW potentials are less than 0.2~kJ/mol.

\begin{table}
\caption{Adsorption energies of different molecules in bulk chabazite obtained for MP2, RPA+GWSE
and two dispersion corrected DFT functionals. Data are in kJ/mol}
\label{tab:ads_dft}
\begin{ruledtabular}
\begin{tabular}{lcccc}
System   & MP2  & RPA+GWSE  & PBE-TS/IH & PBE-D3\\
\hline
Methane  &$-25.6$ &$-26.4$ & $-$31.4  & $-$34.9\\
Ethane   &$-37.0$ &$-38.2$ & $-$47.1  & $-$50.1\\
Ethylene &$-54.2$ &$-52.0$ & $-$56.1  & $-$64.7\\
Acetylene&$-49.1$ &$-46.3$ & $-$50.1  & $-$57.8\\
Propane  &$-47.6$ &$-48.8$ & $-$65.0  & $-$65.4\\
CO$_2$   &$-46.3$ &$-45.6$ & $-$39.8  & $-$48.9\\
H$_2$O   &$-82.7$ &$-82.1$ & $-$83.7  & $-$88.7\\
\end{tabular}
\end{ruledtabular}
\end{table}

\begin{table}
\caption{Adsorption energies of the tested molecules on the 2T model using the 
reference CCSD(T) scheme, RPA+RSE and dispersion corrected DFT. Data are in kJ/mol}
\label{tab:ads_dft_2T}
\begin{ruledtabular}
\begin{tabular}{lcccccc}
System     &CCSD(T)& RPA+RSE& PBE-TS/IH & PBE-D3 & PBE-D4 & PBE0-D4\\
\hline
%Methane    &$-11.0$  &$-$11.1&$-$10.9 & $-$13.0 & $-$12.7 \\
%Ethane     &$-11.6$  &$-$11.8&$-$11.3 & $-$13.2 & $-$12.9 \\
%Ethylene   &$-21.3$  &$-$20.9&$-$22.6 & $-$25.3 & $-$25.4 \\
%Acetylene  &$-16.7$  &$-$16.3&$-$17.0 & $-$20.1 & $-$20.4 \\
%Propane    &$-11.8$  &$-$12.1&$-$12.2 & $-$13.5 & $-$13.1 \\
%CO$_2$     &$-14.6$  &$-$14.2&$-$12.5 & $-$14.3 & $-$14.4 \\
%H$_2$O     &$-44.9$  &$-$43.7&$-$43.7 & $-$46.0 & $-$46.0 \\
Methane    &$-11.0$  &$-$11.1&$-$10.9 & $-$13.0 & $-$13.1& $-$12.6 \\
Ethane     &$-11.6$  &$-$11.8&$-$11.3 & $-$13.2 & $-$13.2& $-$12.7 \\
Ethylene   &$-21.3$  &$-$20.9&$-$22.6 & $-$25.3 & $-$26.0& $-$25.8 \\
Acetylene  &$-16.7$  &$-$16.3&$-$17.0 & $-$20.1 & $-$20.9& $-$20.6 \\
Propane    &$-11.8$  &$-$12.1&$-$12.2 & $-$13.5 & $-$13.3& $-$12.8 \\
CO$_2$     &$-14.6$  &$-$14.2&$-$12.5 & $-$14.3 & $-$14.6& $-$14.7 \\
H$_2$O     &$-44.9$  &$-$43.7&$-$43.7 & $-$46.0 & $-$46.5& $-$47.2 \\
\end{tabular}
\end{ruledtabular}
\end{table}
%& $-$13.4 
%& $-$13.8 
%& $-$25.4 
%& $-$20.1 
%& $-$14.2 
%& $-$14.7 
%& $-$46.7 

%------------------------------------------------------------------------- 
\section{Discussion and Conclusions}
%-------------------------------------------------------------------------

%compare to experiment, finite temperature

We have shown that RPA with singles is highly accurate for predicting adsorption 
energies and we now briefly comment on comparison with experimental data, such as heats of adsorption 
which are available for some of the systems considered here, see, {\it e.g.} Refs.~\onlinecite{hudson2012,pham2013,piccini2015jpcc,hyla2019}
for recent work.
Our interaction energies are obtained for temperature of zero Kelvin and neglecting quantum effects 
of nuclei as well as relaxation of the monomers.
The standard way to approximately treat the first two effects is to obtain zero-point energy
and thermal corrections from vibrational frequencies obtained for molecule bound to the adsorption site.
This approach was used by Piccini~{\it et al.} who notably evaluated the vibrational frequencies 
using anharmonic potential energy surface.\cite{piccini2015jpcc}
However, at room temperature the positions of the molecules are not necessarily limited to a single adsorption site 
and molecular dynamics (MD) is a more appropriate tool to account for the thermal effects.\cite{bucko2011jc}
Given the system sizes and timescales required to converge adsorption energies in zeolites, performing MD based on RPA
is not possible\cite{ramberger2017} and alternative strategies are needed.
Rocca~{\it et al.} have recently used a resampling approach to correct 
PBE-D adsorption enthalpies calculated from MD to RPA level.\cite{rocca2019prm}
Their RPA adsorption enthalpy overestimates the experimental data which is unexpected
given that RPA should underestimate the binding, but we note that it is challenging
to converge the resampling scheme.
To overcome this issue, Chehaibou~and co-workers used machine learning to first train the difference between RPA 
and PBE-D energies and then to use the trained model in the resampling.\cite{chehaibou2019}
This improves the convergence of the resampling significantly and leads to results which are consistent
with the data presented here.
In a similar spirit, Rube\v{s}~{\it et al.} resampled a PBE-based MD with PBE/CC to explain the temperature
dependence of adsorption enthalpy of CO in ferrierite (H-FER).\cite{rubes2018}
The PBE/CC was specifically reparametrized to reproduce
interaction energies at CCSD(T) level for finite clusters and at RPA+RSE level for bulk.
In this work the simulated and measured adsorption heats were within 1~kJ/mol.
Despite these promising results, MD does not account for quantum nuclear effects which could be relevant
even at room temperature for hydrogen-containing molecules.
In any case, RPA with singles can be used to provide reference energies at 0~K with almost reference quality, 
{\it i.e.} with errors of less than 2~kJ/mol or 5~\%.
The values can be further used in resampling of MD or by schemes such as DFT/CC or other\cite{fang2013}
to obtain the finite temperature adsorption enthalpies as shown in Ref.~\onlinecite{rubes2018}, possibly even
with the inclusion of zero point energies, but this is beyond the scope of this paper.

While our focus here was on adsorption energies, we note that RPA with singles was shown to 
give very accurate results for structural properties of solids as well.\cite{klimes2015}
Moreover, Xiao~{\it et al.} found a very good agreement between RPA predictions
and experiment for equilibrium volumes,  bulk moduli, and relative stabilities of
$\alpha$-quartz and stishovite phases of SiO$_2$.\cite{xiao2012}
The accurate description of these properties for zeolites is also problematic for dispersion 
corrected DFT schemes.\cite{fischer2019}
The downside of RPA is its increased computational cost compared to standard DFT
and one can ask to how large systems it can be applied.
Based on our experience, systems with unit cell volumes up to approximately 3000~\AA$^3$
can be studied routinely.
This means that apart from chabazite, zeolites such as ferrierite, mordenite, or LTA are accessible. 
As RPA does not rely on any parameters, it can be also readily applied to different cation-exchanged
zeolites which show interesting adsorption properties and for which reference data would 
helpful.\cite{zhang2008,denayer2008,shang2012,shang2014,coudert2017}
Finally, RPA with singles can also provide adsorption energies and structural properties of nearly reference
quality to other porous materials, such as MOFs and ZIFs.\cite{wieme2018}

In summary, we have shown that the RPA with singles corrections scheme surpasses
the MP2 approach for obtaining adsorption energies of molecules in zeolites.
This is not only in terms of accuracy but also in terms of computational cost.
We demonstrated this on a test set containing seven diverse molecules.
The lower reliability of MP2 was the most apparent for ethylene and acetylene.
These molecules possess delocalized electrons for which the description of electron 
screening is needed, this is present in RPA, but not in MP2.
Overall, RPA with singles is a suitable scheme for obtaining highly accurate 
adsorption energies that can be used to compare to experimental data and to assess
or even enhance the accuracy of simpler theoretical methods.

%-----------------------------------------------------------------------------------------
\begin{acknowledgements}
%-----------------------------------------------------------------------------------------
JK acknowledges support from the European Union's Horizon 2020 research and innovation
programme under the Marie Sklodowska-Curie grant agreement No 658705
and via ERC grant APES (No 759721) and support by a PRIMUS project from Charles University.
DPT acknowledges support from the Max Planck Institute.
Computational resources were provided by the MetaCentrum and CESNET (LM2015042),
CERIT-SC (LM2015085), and by IT4Innovations (LM2015070), supported by the Czech Ministry of Education, 
Youth, and Sports.
We thank Florian G\"{o}ltl for providing us with some of the initial structures.
We thank Miroslav Rube\v{s} for useful discussions.
\end{acknowledgements}

%\begin{suppinfo}
%\begin{itemize}
%  \item structures.tgz: The structures of the systems studied.
%  \item inputs.tgz: Input for VASP used to obtain the energies.
%\end{itemize}
%\end{suppinfo}

%-----------------------------------------------------------------------------------------
%\section*{References}
%\bibliography{TS}
%-----------------------------------------------------------------------------------------

\end{document}